\documentclass[letterpaper, 10 pt, conference]{ieeeconf}  

\IEEEoverridecommandlockouts                              

\usepackage{times}
\usepackage{epsfig}
\usepackage{graphicx}
\usepackage{amsmath}
\usepackage{hyperref}
\usepackage{cite}

\usepackage{subcaption}
\DeclareCaptionSubType*[Alph]{table}
\DeclareCaptionLabelFormat{mystyle}{Table~\bothIfFirst{#1}{ ̃}#2}
\usepackage{multirow}
\usepackage{array}
\newcolumntype{P}[1]{>{\centering\arraybackslash}p{#1}}

\usepackage{xcolor}
\usepackage{comment}

\title{\LARGE \bf
If you Cheat, I Cheat: Cheating on a Collaborative Task with a Social Robot
}

\author{Ali Ayub$^{1}$, Huiqing Hu$^{2}$, Guangwei Zhou$^{3}$, Carter Fendley$^{4}$, Crystal Ramsay$^{5}$,\\ Kathy Lou Jackson$^{6,a}$ and Alan R. Wagner$^{6,b}$
\thanks{$^{1}$Department of Electrical Engineering,
        The Pennsylvania State University, State College, PA 16802, USA
        {\tt\small aja5755@psu.edu}}%
\thanks{$^{2}$Outreach,
        The Pennsylvania State University, State College, PA 16802, USA
        {\tt\small huh162@psu.edu}}
\thanks{$^{3}$Department of Robotics, 
        Johns Hopkins University, Baltimore, MD, 21218, USA
        {\tt\small gzhou11@jhu.edu}}
\thanks{$^{4}$Department of Computer Science, 
        The Pennsylvania State University, State College, PA 16802, USA
        {\tt\small ccf5164@jhu.edu}}
\thanks{$^{5}$Teaching and Learning with Technology, 
        The Pennsylvania State University, State College, PA 16802, USA
        {\tt\small cmg5@psu.edu}}
\thanks{$^{6}$Department of Aerospace Engineering, The Pennsylvania State University,
        State College, PA 16802, USA
        {$^{a}$\tt\small klj11@psu.edu}, $^{b}$ \tt\small alan.r.wagner@psu.edu}
}

\begin{document}

\maketitle
\thispagestyle{empty}
\pagestyle{empty}

\begin{abstract}
\label{sec:Abstract}
Robots may soon play a role in higher education by augmenting learning environments and managing interactions between instructors and learners. Little, however, is known about how the presence of robots in the learning environment will influence academic integrity. This study therefore investigates if and how college students cheat while engaged in a collaborative sorting task with a robot. We employed a 2x2 factorial design to examine the effects of cheating exposure (exposure to cheating or no exposure) and task clarity (clear or vague rules) on college student cheating behaviors while interacting with a robot. Our study finds that prior exposure to cheating on the task significantly increases the likelihood of cheating. Yet, the tendency to cheat was not impacted by the clarity of the task rules. These results suggest that normative behavior by classmates may strongly influence the decision to cheat while engaged in an instructional experience with a robot.

\end{abstract}
\section{Introduction}
\label{sec:introduction}
\noindent Robots may soon play a role in higher education by augmenting learning environments and managing interactions between instructors and learners. Yet, very little is known about how the presence and use of robots in higher education will impact academic integrity. There are reasons to be concerned. McCabe, Butterfield, and Travino note that cheating is widespread and increasing \cite{mccabe2012cheating}. Davis, Drinan, and Gallant suggest that methods for cheating have evolved with technology \cite{davis_drinan_gallant_2011}. Given the prevalence of cheating and the possibility that technology makes it easier to cheat, we set out to investigate how a new and not yet common technology in education, robots, might persuade or dissuade students from cheating. 

This exploration is important because robots are currently being developed to serve as an educational tool \cite{belpaeme2018social}. These robots may operate in a variety of different roles ranging from tutoring \cite{kose2015effect}, acting as a teaching assistant \cite{obaid2015designing}, or as a novice that a student must teach \cite{hood2015children}. The development and introduction of robots into the classroom is motivated by the increasing number of students per classroom, the desire to meet diverse educational needs, and the view that new technologies benefit students. Admittedly, the use of robots in the classroom is largely limited to research under well-controlled environments \cite{belpaeme2018social}. Nevertheless, it is useful to explore the possible ramifications of a technology before the technology becomes widespread. Doing so may shed light on some of the technology’s downsides thus allowing educators to better assess whether the investment is worth the cost.

The purpose of this study is to investigate if and how college students cheat when engaged in a collaborative task with a robot, with a particular focus on the roles of exposure to peer cheating and task clarity. We choose this experimental setup to model academic situations in which the student may decide to violate the assignment’s rules in order to obtain high grades on the assignment when collaborating with a robot. Our investigation focused on whether or not college students would cheat while teaming with a robot on a task that they could benefit from academically and, if so, to explore the circumstances under which they were most likely to cheat. We employed a 2x2 factorial design, with the first factor being cheating exposure through peer behaviors (i.e., confederate cheated or did not cheat) and the second factor being the clarity of the task rules (i.e. vague rules or clear rules). We made the following two hypotheses:
\begin{enumerate}
    \item Students would be more likely to cheat after they observe another student cheat on the same task.
    \item Students in the vague rules condition would be more likely to cheat compared to those who received a clear set of the task rules.
\end{enumerate}


The remainder of this paper begins with an examination of the related work. This is followed by a description of the experimental methods used, followed by the experimental results. The paper concludes by considering the studies limitation and assumptions and discussing avenues for future research.

\section{Related Work}
\label{sec:related_work}
\noindent Academic integrity is central to an ethical learning environment. However, academic dishonestly is widespread and increasing \cite{davis2009cheating}. Contextual factors play a critical role in motivating students to cheat \cite{mccabe2012cheating}. For instance, observing a peer cheat, and presenting vague rules for a certain task has been linked to greater tendencies for cheating behaviors in academic settings \cite{carrell_malmstrom_west_2008,mccabe2012cheating}. 



Research on the impact of integrating robots in academic contexts is in its infancy. Research has examined how the presence of a robot influences ethical behavior \cite{forlizzi2016let}. Forlizzi et al. \cite{forlizzi2016let} employed a controlled field study in which unsuspecting subjects passed by food labeled 'reserved' with no observer, a human, or a robot watching the food. They found that people behave more dishonestly when the food is being monitored by the robot than the human. Relatedly, Petisca et al. investigate whether or not people are more likely to cheat at a game while in the presence of a robot. They find that the presence of the robot does not inhibit cheating, and find that, in fact, the person cheats at the same rate as if they were alone \cite{petisca2019cheating,petisca2020_ACM}. Both of these works considered cheating scenarios in which the robot was merely an observer and did not take part in the task. 

This work investigates a gap in our understanding related to how student academic integrity could be impacted during an instructional experience with a robot. To the best of our knowledge, this is the first paper to examine the factors that influence students to cheat while collaborating with a robot.



\section{Method}
\label{sec:method}
\subsection{Participants}
\label{sec:participants_info}
\noindent We recruited 78 college students (18-22 years old) from three introductory educational psychology courses at The Pennsylvania State University. Most participants were female (78\%), freshman (62\%), and self-reported their GPA for the current semester as between 3.6-4.0 (36\%). The experimental conditions did not differ significantly by gender, $\mathcal{X}^2(3)=1.21$, $p=0.75$, or self-reported GPA, $\mathcal{X}^2(6)=8.02$, $p=0.24$. Our study was IRB approved by the Penn State office of research.  

\begin{figure}[t]
\centering
\includegraphics[width=1.0\linewidth]{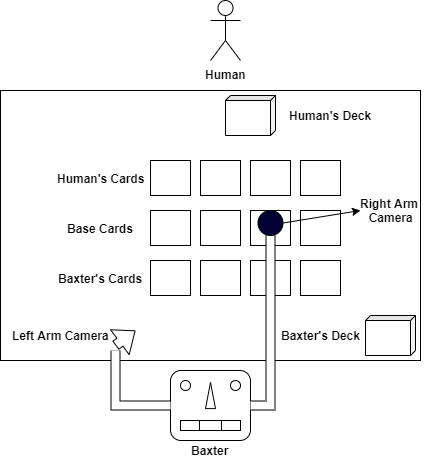}
\caption{\small Experiment layout depicting the participant, the Baxter robot, and the cards needed for the task. The human and the robot were on different sides of the table. Base cards were placed on the table with specified slots for the human's and the robot's cards. The robot's right arm hovered over the cards to get a top view of the table and moved to pick and place cards. The robot’s left arm was static and was used to record video of the experiment.}
\label{fig:setup}
\end{figure}

\begin{figure}[t]
\centering
\includegraphics[width=1.0\linewidth]{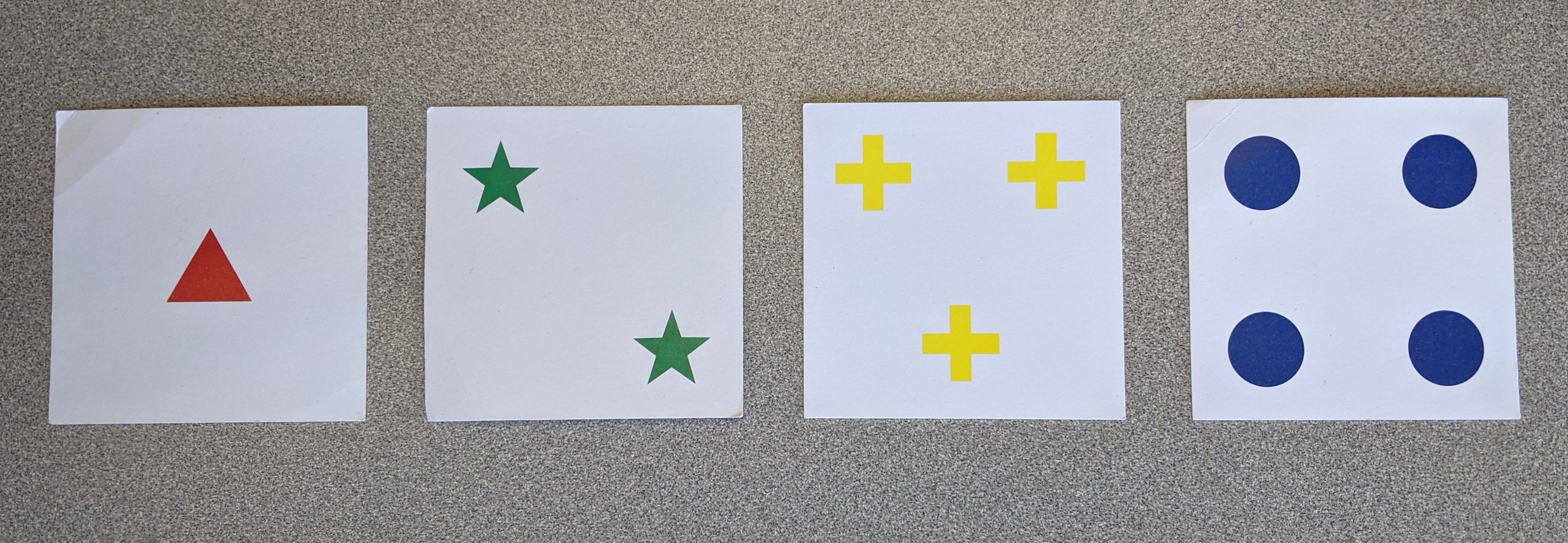}
\caption{\small The four baseline Wisconsin cards used in our experiment.}
\label{fig:wisconsin_cards}
\end{figure}

\subsection{The Collaborative Task}
\label{sec:task_info}
\noindent We adapted the Wisconsin Card Sorting Task (WCST) to serve as the collaborative task for this study \cite{kopp2019}. In the original WCST, each card is characterized by a shape (stars, circles, triangles, squares), color for the shapes (yellow, green, blue, or red), the number of shapes (1 - 4). Four baseline WCST cards are present in front of the participant: a card with one red triangle, a card with two green stars, a card with three yellow crosses, and a card with four blue circles (see Figure \ref{fig:wisconsin_cards}). During each turn, the participant sorts a card by placing it in front of a baseline card based on one of the three characteristics, known as the sorting criterion. After the sort, the participant receives a feedback sound indicating whether or not the sort was correct based on the current sorting criterion. The participant is not told the sorting criteria during the task, rather he/she has to use the feedback after the sort to deduce the sorting criteria. After a small number of sorts, the sorting criterion changes randomly and the participant has to learn the new criterion.  

For this study, we converted the original WCST into a collaborative task, where the human participant and the Baxter robot took turns sorting the Wisconsin cards, instead of sorting individually. The robot was situated on one side of the experiment table; the participants sat on the other side of the table upon arrival. The task environment included four baseline WCST cards which were placed on the table, along with a timer, a speaker, and two decks of additional cards. 
See Figure \ref{fig:setup} for the experimental layout.

Before the task began, participants were told to choose any of three criteria to begin sorting (e.g., color), and the robot would sort next. They were also told the correct sorting criterion would change throughout the task, but that participants would not be given the criterion ahead of time. They were told to pay attention to the feedback, a sound from the an external speaker after each sort, to deduce the criterion. A “ding” sound indicated that a card was sorted correctly, and a “buzz” sound indicated that the card was not sorted correctly. The researcher controlled the speaker and played feedback sounds after each sort made by the participant or the robot. The researcher predetermined the pattern of the sorting criteria (e.g., started with color, then changed to number, then a shape). The criteria switch after 5–7 sorts. On average, the sorting criteria changed 3–6 times for each participant over the course of the experiment.  

Participants were instructed to take turns with the robot. Participants were also instructed that if they sorted 10 cards correctly within the 7 minutes allotted, and if Baxter also sorted 10 cards correctly, they would receive extra credit (course points) toward their course grade. 
Otherwise, if they and/or Baxter fell short of these goals then they would not receive extra credit. However, this instruction was only used as a motivation.  All the participants were awarded extra credit regardless of whether or not they sorted the required number of cards. 

Across all conditions, the robot sorted 8-9 cards at a slow, steady pace. The robot always sorted below the threshold needed to succeed at the task (i.e., 10 cards). In addition, some cards were purposefully sorted incorrectly in order to represent a poor performing partner. The robot’s behavior was designed to put the extra credit in jeopardy motivating the participants to consider cheating on the task.

\begin{figure}[t]
\centering
\includegraphics[width=1.0\linewidth]{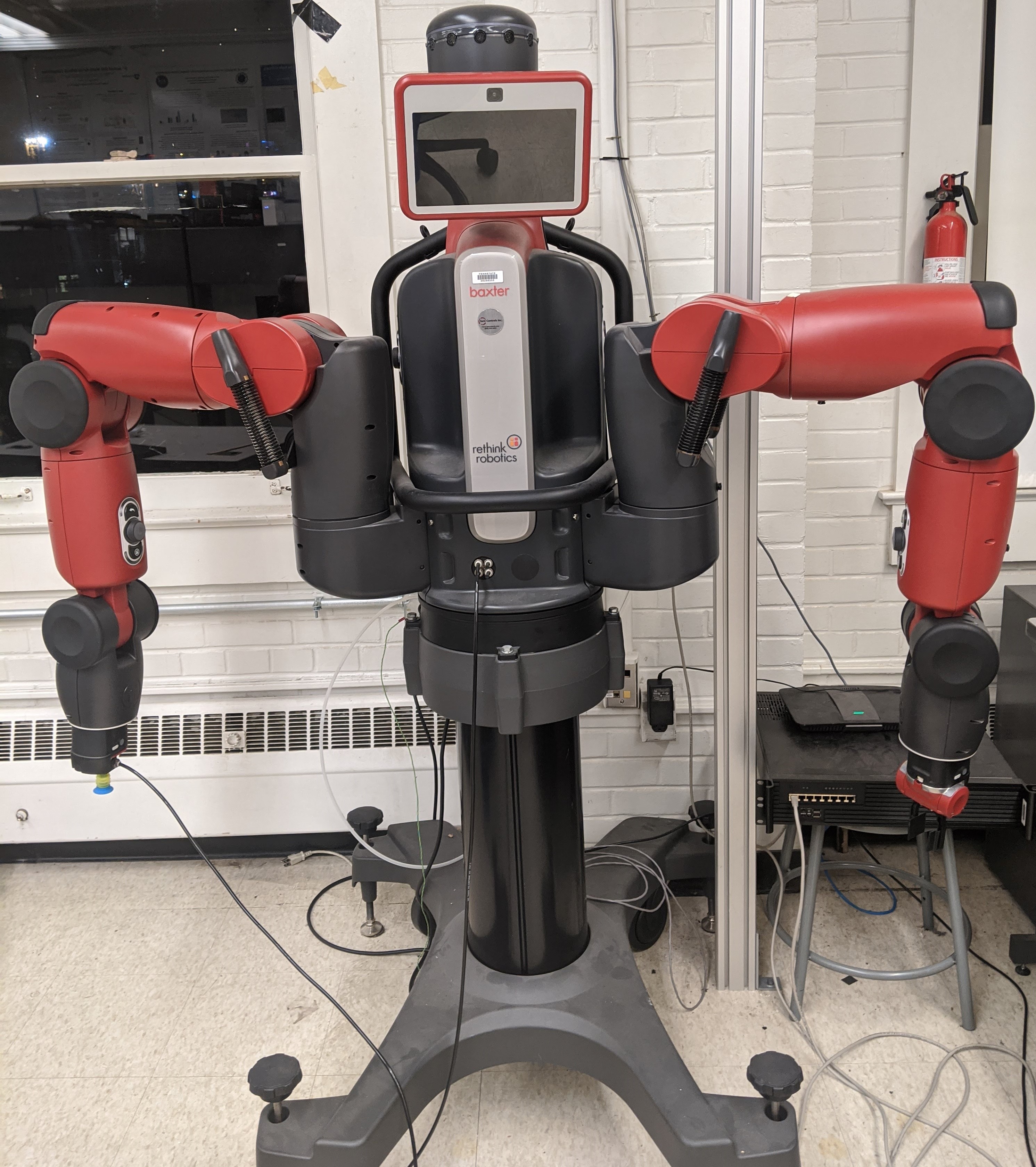}
\caption{\small The Baxter robot.}
\label{fig:baxter}
\end{figure}

\subsection{The Baxter Robot}
\noindent The Baxter robot, manufactured by Rethink Robotics, was used for this experiment (see Figure \ref{fig:baxter}).  
The suction gripper on the robot's right arm was used to pick up and move cards on a flat table. Baxter has three cameras, one on its head and two cameras near its hands. The hand cameras were used to take images and record videos while performing tasks. The robot was controlled remotely by the researcher although human participants believed that the robot was acting autonomously. The robot was programmed to pick up cards from a pre-defined location on the table and to move cards to one of four pre-defined locations on the table. The robot’s right arm was used to move the cards. Its arm hovered over the table when not moving allowing the hand camera to capture video of the table surface. The camera output was sent to a remote desktop allowing the researcher to monitor the table and the cards. The robot’s left arm remained stationary with its camera pointing towards the table as well (Figure \ref{fig:setup}). The camera output from the left arm was also sent to the remote desktop providing a second view of the table.

During the robot's turn to sort a card, the researcher first commanded the robot to pick up a card by pressing a button on the keyboard. After the robot picked up the card using its suction gripper, the arm moved back to the default position over the table. Next, the researcher chose one of the four places (for four base cards) on the table by typing a number between (1-4) on the keyboard. The robot then placed the card at the specified location. 

\begin{table*}
\centering
\small
\begin{tabular}{|P{4.0cm}|P{8.5cm}|P{3.0cm}|}
     \hline
    \textbf{Type of Cheating Behavior} & \textbf{Description} & \textbf{Demonstrated by}\\
     \hline
     Look Ahead & After the robot’s turn, the confederate flipped the next card over and looked at it to prepare for the next turn. & Confederate and participant \\
     \hline
     Did not wait for the robot's turn & Before the robot starts to move for its next turn, the confederate sorted another card into one of his sorted card slots with either one of his or Baxter’s card. & Confederate and participant \\
     \hline
     Sorted for the robot & The confederate put a card in one of the robot’s sorted cards slots with either the robot’s or his own card. & Confederate and participant \\
     \hline
     Sort the same card again & Immediately after the participant gets feedback on a sort, they sort the same card again in another one of their sorted card slots. & Participant \\
     \hline
     Sort for the robot if it fails to pick up & The participant sorts for the robot when the robot fails to pick up the card due to a suction gripper failure. & Participant \\
 \hline
 \end{tabular}
 \caption{\small Descriptions of the types of cheating behaviors occurring in the study.}
 \label{tab:cheating_behaviors}
 \end{table*}

\subsection{Confederates}
\label{sec:confederates}
Two confederates were hired for this experiment to act as a participant and to demonstrate cheating behaviors in the confederate cheating condition (explained in Section \ref{sec:experiment_conditions}). 
The first confederate was a male undergraduate student, 21 years old, Caucasian, a junior, and majoring in computer science. The second confederate was also a male undergraduate student, 21 years old, Asian, a junior, and majoring in mechanical engineering.

\subsection{Experimental Conditions}
\label{sec:experiment_conditions}
Participants were randomly assigned to one of the four conditions: confederate cheating, vague rules (CV); confederate cheating, clear rules (CC); no confederate cheating, vague rules (NV); and no confederate cheating, clear rules (NC). 

In the cheating exposure conditions (CV \& CC), a confederate used three different cheating behaviors: looked ahead, did not wait for the robot’s turn, and sorted for the robot. See Table 2 for descriptions of each cheating behavior. Both confederates demonstrated approximately the same number of cheating behaviors. They also expressed frustration before and after they cheated by sighing, shaking their head, and/or showing anxious facial expressions. In the non-cheating conditions (i.e., NV \& NC), confederates did not display any cheating behaviors or express frustration. Instead, confederates followed all of the rules for the task. 

In the clear rule conditions (i.e., CC \& NC), the researcher verbally explained the rules and provided a list of the rules for the sorting task that was displayed on the table throughout the experiment. The participants in this condition were informed that the list of the rules on the table were the same as the ones explained to them, and they could review the rules at any time during the experiment. In the vague rule conditions (i.e., CV \& NV), the researcher verbally explained the rules of the sorting task to each participant before the start of the experiment but the list of rules were not displayed on the table. Participants were also not given time to ask questions.


\subsection{Procedures}
\label{sec:Procedures}
\noindent The experiment was conducted in a robotics laboratory. For each session, one of the confederates arrived at the lab at approximately the same time as the participant and pretended to be a participant for the same experiment. When both the participant and the confederate arrived, the researcher greeted them, asked for their names, explained that there was only one set up for the study so the two participants (the participant and the confederate) would have to take turns completing the experiment. The confederate was asked to go first, followed by the real participant. The confederate was directed to complete a pre-experiment Qualtrics survey that was loaded on a laptop placed next to the experiment table, and the participant was asked to wait for their turn. Once the survey was complete, the confederate was asked to sit in front of the experiment table and the participant was instructed to complete the survey. Once the participant completed the survey, the experiment began with the researcher verbally explaining the instructions for the sorting task to both individuals. 

The confederate and the participant were then introduced to the Baxter robot. They were shown each part of the task and the researcher explained the rules of the sorting task. The participant was told that the first “participant” (i.e., the confederate) would complete the sorting task with Baxter first, then it would be the participant’s turn. Once the instructions were given, the researcher started the timer remotely for the confederate to begin his turn. The participant was able to clearly observe the confederate, their sorting behavior, and the whole experiment table while waiting for his or her turn. The researcher did not stay in the experiment area when the confederate or the participant were completing the task.

\begin{figure}[t]
\centering
\includegraphics[width=1.0\linewidth]{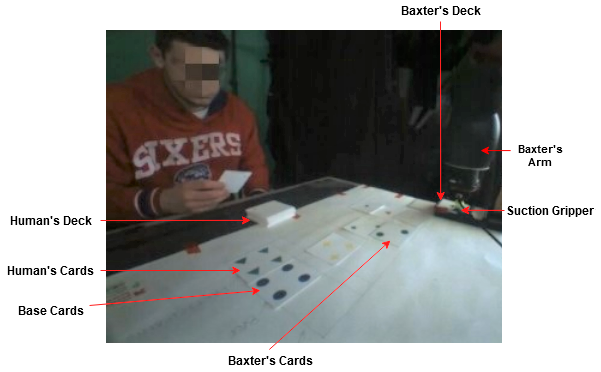}
\caption{\small Image depicting the layout of the cheating experiment. This image was artificially brightened to increase readability.}
\label{fig:paricipant_example}
\end{figure}

After the confederate’s turn, the researcher stopped the timer, and told the confederate “Congratulations, you will be getting the extra credit, because you got more than 10 of your cards sorted correctly and Baxter got more than 10 of his cards sorted correctly.” This statement was made in all conditions regardless of how many correct cards they sorted. The confederate was then asked to switch seats with the participant. The participant completed the same sorting task while the confederate completed the post-experiment survey. Figure \ref{fig:paricipant_example} shows an example of a participant completing the WCST task. Immediately after the participant completed the sorting task, they were instructed to complete the post-experiment survey.

At the end of the experiment, participants were fully debriefed. They were told that they would receive the extra credit regardless of how many cards they and the robot sorted correctly, and that the first “participant” was a confederate. They were then offered an opportunity to ask the researcher questions about the nature of the experiment. All the sessions were video recorded for data analysis purposes and all the participants provided permission to use their video data.

\begin{table*}
\centering
\small
\begin{tabular}{P{4.0cm}P{1.0cm}P{1.0cm}P{1.0cm}P{1.0cm}P{1.0cm}P{1.0cm}P{1.0cm}P{1.0cm}}
     \hline
     \multicolumn{1}{c}{} & \multicolumn{2}{c}{\textbf{CV}} & \multicolumn{2}{c}{\textbf{CC}} & \multicolumn{2}{c}{\textbf{NV}} & \multicolumn{2}{c}{\textbf{NC}}\\
     \hline
     & \textbf{M} & \textbf{SD} & \textbf{M} & \textbf{SD} & \textbf{M} & \textbf{SD} & \textbf{M} & \textbf{SD}\\
    \hline
     Anthropomorphism & 2.26 & 0.80 & 2.37 & 0.72 & 2.40 & 0.73 & 2.34 & 0.56 \\
     Likeability & 3.50 & 0.55 & 3.48 & 0.72 & 3.48 & 0.65 & 3.33 & 0.50 \\
     Perceived Intelligence & 3.95 & 0.59 & 4.14 & 0.67 & 3.86 & 0.58 & 3.72 & 0.47 \\
 \hline
 \hline
    & \textbf{No.} & \textbf{\%} & \textbf{No.} & \textbf{\%} & \textbf{No.} & \textbf{\%} & \textbf{No.} & \textbf{\%}\\
     \hline
     Prior interaction with a robot & 4 & 19 & 2 & 10 & 5 & 10 & 2 & 10 \\
     Prior exposure to WCST & 1 & 5 & 2 & 10 & 1 & 6 & 2 & 10 \\
     Prior completion of WCST & 0 & N/A & 1 & 6 & 1 & 6 & 1 & 5 \\
 \hline
 \end{tabular}
 \caption{\small Means and standard deviations for the four conditions on participants’ impressions of robots (Top). The number and percentage of participants with prior exposure to robots and the WCST (Bottom).}
 \label{tab:godspeed_pre_survey}
 \end{table*}

\subsection{Measures}
\label{sec:measures}
\noindent Several different metrics were recorded in order to capture the participant's cheating behaviors and opinions about the task. 
\subsubsection{Type and Frequency of Cheating Behaviors}
\label{sec:freq_behaviors}
The type and frequency of participants’ cheating behaviors were coded using the video recordings of each participant. We used a qualitative research tool, MAXQDA 2018\footnote{Software available from \url{maxqda.com}} for the video analysis. To establish the codes, two coders applied a general inductive approach to examine the types of cheating behaviors displayed by the participants, without imposing a preexisting framework for the behaviors that were expected to be observed \cite{thomas2006}. Specifically, the two coders independently watched video recordings of the cheating conditions and each created an initial list of distinct cheating behaviors. The two coders then combined their lists of initial cheating behaviors, further refined the combined list through discussion, and came to consensus on a final list of five unique cheating behaviors. As indicated in Table \ref{tab:cheating_behaviors}, the five types of cheating behaviors included: (1) look ahead, (2) does not wait for the robot’s turn, (3) sort for the robot, (4) sort the same card again and (5) sort for the robot if it misses. The first three behaviors were demonstrated by the confederate and imitated by the participants, the latter two were new cheating behaviors demonstrated by the participants during the study. Based on the final coding scheme, the two coders independently analyzed the video data for the cheating conditions and coded the frequencies of each cheating behavior. Intra-class correlation using a two-way mixed model with absolute agreement was conducted on the two raters’ coded frequencies of each cheating behavior, and interrater reliability was lowest for "sort for the robot if it misses" [ICC (3,2) = .444] and highest for "sort for the robot" [ICC (3,2) = .968]. Thereafter, the two raters revisited the coding scheme and resolved all the frequency discrepancies in the cheating conditions and reached perfect agreement. Finally, each rater coded the remainder of the videos, reconciled, and resolved all of the remaining frequency disagreements. The two coders agreed on all the coded behaviors in the final version of the data set. 

\subsubsection{Pretest and Posttest Measures}
\label{sec:pretest_measures}
Participants’ existing perception of robots was measured at the beginning of the experiment using the anthropomorphism (5-items), likeability (5-items), and perceived intelligence (5-items) subscales on the Godspeed questionnaire \cite{bartneck2008,weiss2015}. Each item had a unique 5-point semantic differential scale with bipolar anchors (e.g. 1=Fake, 5=Natural; 1=Machinelike, 5=humanlike), and participants were asked to rate their impressions of robots on a total of 15 items across the three subscales. These subscales have been shown to have strong reliability with Cronbach’s $\alpha$ over .70 \cite{ho2010}. Likewise, the current study also found high internal consistency among all three dimensions: anthropomorphism (Cronbach’s $\alpha$=0.7), likeability (Cronbach’s $\alpha$=0.85), and perceived intelligence (Cronbach’s $\alpha$=0.77).

On the posttest, two items were used to measure participants’ prior experiences with robots. Participants were asked if they had ever interacted with a robot (yes/no) prior to the experiment. If they selected yes, they were then asked to describe their prior interactions with a robot in an open-ended format. Next, two more items were used to measure participants’ prior experiences with the Wisconsin Card Sorting Task (WCST). Participants were asked if they had heard of the WCST (yes/no). If they responded yes, they were then asked if they have experience completing the WCST prior to the experiment (yes/no). The 9-item Academic Dishonesty Scale \cite{mccabe1993academic} was given to measure self-reported cheating behaviors in academic settings. On a 5-point Likert scale (1 = never, 5 = many times), participants were asked to rate the frequency with which they had engaged in nine common academically dishonest behaviors (e.g. “Copying a few sentences of material without footnoting in a paper”). Prior studies have reported the scores on this scale with high reliability (Cronbach’s $\alpha$=0.79) \cite{mccabe1993academic}. Consistent with prior reporting, scores on the scale in the present study were also found to be internally reliable (Cronbach’s $\alpha$=0.81). Finally, participants’ perception of their own cheating behaviors was measured with a yes/no question asking if they thought they had cheated during the collaborative task with the robot. If they selected yes, they were then asked in an open-ended format, how they cheated and why they cheated.

\section{Results}
\label{sec:results}

\begin{table*}
\centering
\small
\begin{tabular}{P{2.0cm}P{1.0cm}P{1.0cm}P{1.0cm}P{1.0cm}P{1.0cm}P{1.0cm}P{1.0cm}P{1.0cm}P{1.0cm}}
     \hline
     \multicolumn{1}{c}{} & \multicolumn{3}{c}{\textbf{Exposed to Cheating}} & \multicolumn{3}{c}{\textbf{Not Exposed to Cheating}} & \multicolumn{3}{c}{\textbf{Total}}\\
     \hline
     & \textbf{n} & \textbf{M} & \textbf{SD} & \textbf{n} & \textbf{M} & \textbf{SD} & \textbf{n} & \textbf{M} & \textbf{SD}\\
    \hline
     Vague Rules & 21 & 15.29 & 10.47 & 17 & .76 & 2.11 & 38 & 8.03 & 10.71 \\
     Clear Rules & 20 & 11.50 & 9.09 & 20 & 1.05 & 2.09 & 40 & 6.28	& 8.39 \\
     Total & 41	& 13.40 & 1.14 & 37 & .91 & 1.20 & 78 & 7.50 & 9.61 \\
 \hline
 \end{tabular}
 \caption{\small Means and standard deviations of the average occurrences of cheating behaviors by condition and the total number of participants.}
 \label{tab:peer_cheating}
 \end{table*}

\subsection{Preliminary Analysis}
\label{sec:preliminary_analysis}
\noindent A preliminary analysis was conducted to examine whether the experimental conditions were balanced based on participants’ prior exposure to robots and the WCST. Descriptive analysis is shown in Table \ref{tab:godspeed_pre_survey}. Three one-way ANOVAs on each of the Godspeed subscales showed that there were no statistically significant differences across the conditions on participants’ impressions of robots in terms of anthropomorphism, $F(1,74) = .09$, $p=.97$; likeability, $F(1,74) = .33$, $p=.80$; or perceived intelligence, $F(1,74) = 1.82$, $p=.15$. Few participants (10\% in CC \& NC, 19\% in CV, 29\% in NV)  
reported that they had interacted with a robot such as Baxter robot prior to the experiment. A chi-square test indicated that there were no statistically significant differences across the conditions in the proportion of participants who had this experience, $\mathcal{X}^2 (3)=3.35$, $p=.34$. Among the participants who had experiences with robots, none indicated that they engaged in a sorting task with a robot or had collaborated with the Baxter robot, in particular, before. Moreover, very few participants indicated that they had heard of the WSCT (5\% in CV \& NC, 6\% in NV, 10\% in CC) 
or had completed the task (0 in CV, 5\% in NC, 6\% in CC and NV) 
prior to the study. Separate chi-square tests showed there were no statistically significant differences across the conditions among those who had heard about the WSCT, $\mathcal{X}^2 (3)=.60$, $p=.90$, or had completed the task, $\mathcal{X}^2 (3)=1.18$, $p=.76$. In sum, the four 
experimental conditions were not statistically different with respect to participants' impressions of robots, prior interactions with robots, and prior exposure to the WCST.

\begin{table}[t]
\centering
\small
\begin{tabular}{|P{2.5cm}|P{0.9cm}|P{0.9cm}|P{0.9cm}|P{0.9cm}|}
     \hline
    \textbf{Participants} & \textbf{CV} & \textbf{CC} & \textbf{NV} & \textbf{NC}\\
     \hline
     who were observed cheating & 20 (95\%) & 16 (80\%) & 2 (12\%) & 5 (25\%) \\
     \hline
     who reported that they cheated & 13 (62\%) & 9 (45\%) & 1 (6\%) & 1 (5\%) \\
     \hline
     Total & 21 & 20 & 17 & 20 \\
 \hline
 \end{tabular}
 \caption{\small The number and percentage of participants who were observed cheating and who self-reported that they cheated in the four conditions. Total number of participants in each condition are also reported.}
 \label{tab:do_students_cheat}
 \end{table}

\subsection{Do students cheat when teamed with a robot?}
\label{sec:student_cheat_experiment}
As stated in Section \ref{sec:related_work}, one unique aspect of our experimental setup was that the robot acted as a peer collaborator with a participant rather than an observer or an invigilator. The first question we hoped to answer was whether college students would cheat on a collaborative task with a robot if their reward was contingent on completing the task and the robot was a poor partner.
Table \ref{tab:do_students_cheat} presents the number of participants who cheated during the sorting task with the robot, based on the observations of the researchers and the number of participants who self-reported that they had cheated on the task. 
In our study, we found that instances of cheating ranged from a minimum of two people (12\%) in the NV condition to a maximum of 20 people (95\%) in the CV condition. Interestingly, not all students who cheated believed that they had cheated. Overall, approximately half of the participants self-reported that they followed the task rules. 
In the CV condition, 65\% of participants (13 individuals) who cheated reported that they cheated, 56\% (9 people) of those who cheated in CC reported that they cheated, and 50\% (1 person) of those who cheated in the NV condition reported that they cheated. In the NC condition, only one out of the five participants who cheated self-reported that they broke the rules. 
These results suggest that a large portion of students that cheated did not report their behavior as cheating. Overall, these findings indicate college students cheated on the sorting task while collaborating with the robot and that they were sufficiently motivated by the extra credit offered to do so.

\subsection{Do students cheat more after being exposed to a peer’s cheating? Do students cheat more when given vague rules?} 
\label{sec:confederate_cheating_experiment_results}

\noindent 
Table \ref{tab:peer_cheating} presents the means and standard deviations of the average occurrences of cheating behaviors by cheating exposure and clarity of rules. A two-way ANOVA was conducted on the number of cheating behaviors, with cheating exposure (exposed or not exposed to confederate cheating) and rule clarity (vague or clear rules) serving as the between-subject factor. The analysis indicated that there was a statistically significant main effect for cheating exposure, $F(1,78)=57.03$, $p<.001$, partial $\eta^2=0.44$., but not rule clarity,$F(1,78)=1.12$, $p=.29$. Pairwise comparisons with Bonferroni adjustments showed that participants who were exposed to the confederate cheating performed significantly more cheating behaviors (estimated marginal mean= 13.34) than those who did not observe the confederate cheat (estimated marginal mean=.91; $M_{difference}=12.49$)., regardless of the clarity of the rules. No statistically significant interaction effect between the two factors was found, $F(1,78)=1.51$, $p=.22$.  

These results show that college students who observed a peer cheating on a task when collaborating with a robot were significantly more likely to cheat on the task themselves. Further, students’ cheating behaviors were not influenced by the vagueness of the task rules and the effects of cheating exposure on cheating behaviors were not dependent on rule clarity.


\section{Conclusion}
\label{sec:conclusion}
\noindent This research has taken a novel and important approach to investigating how the presence of robots in the learning environment could influence academic integrity. We have shown that, when tasked with collaborating with a poorly performing robot and motivated to meet the task demands, students will cheat to complete the task only if they have witnessed a peer cheat. This is an important result that sheds light both on the conditions that underpin academic dishonesty and a possible prelude to the issues generated by placing robots in educational environments. Broadly construed, these results suggest that normative behavior by classmates may strongly influence the decision to cheat while engaged in an instructional experience with a robot. 

Our experiment did not compare collaboration with a robot to collaboration with a person. We felt that the participant would have encouraged and impelled a poor performing human partner to do better. Using the robot allowed us to generate a situation in which the person had to either accept the poor performance or cheat. Our data shows that for most, cheating only became an option once they witnessed the confederate cheating. 

The results from this research may extend beyond educational environments. Although the research was experimentally framed as an educational task, we conjecture that these results may apply to other applications such as healthcare. For example, it may be the case that when a patient is tasked with collaborating with a robot to complete rehabilitative exercises the patient is more inclined to cheat on those exercises. Whether or not this is the case remains to be seen.





\section*{Acknowledgment}
\noindent Support for this research was provided by a grant from Penn State's Teaching and Learning with Technology (TLT) Center.


{\small
\bibliographystyle{IEEEtran.bst}
\bibliography{main}
}

\end{document}